# New Model of Network- A Future Aspect of the Computer Networks

Ram Kumar Singh, Prof. T.Ramajujam

**Abstract**— As the number and size of the Network increases, the deficiencies persist, including network security problems. But there is no shortage of technologies offered as universal remedy – EIGRP,BGP, OSPF, VoIP, IPv6, IPTV, MPLS, WiFi, to name a few. There are multiple factors for the current situation. Now a day during emergent and blossoming stages of network development is no longer sufficient when the networks are mature and have become everyday tool for social and business interactions. A new model of network is necessary to find solutions for today's pressing problems, especially those related to network security. In this paper out factors leading to current stagnation discusses critical assumptions behind current networks, how many of them are no longer valid and have become barriers for implementing real solutions. The paper concludes by offering new directions for future needs and solving current challenges.

——————————— ◆ ———————————

## 1 Introduction

THE current remorseful state of networking can be realized by reflecting on this scenario: Voice over internet protocols (VOIP) and Email are the most popular network (Internet) application today. But email spam, essentially a technical issue requiring a technology solution, is a plague affecting all network users.

Transfer Network Architecture (TNA) is a new model created by superimposing an architecture framework to the existing network infrastructure [7]; consisting of the Internet, Signaling System 7 (SS7) network, data circuits of the PSTN, and other networks [11]. The current packet-centric strategies are not helpful to meet the network needs of today and tomorrow. Superior network capabilities will emerge by promoting technology and protocol agnostic inter-operating networks, with each network implemented by optimizing its design requirements derived from service needs, without extraneous considerations. The Transfer Network Architecture (TNA) is a reference model that can help such developments.

────────────────────────────
- **Ram Kumar Singh** *is with the Department of Information Technology, International School of Business & Tech. Kampala, Uganda, East Africa. Post Box-28220*

- **Prof. T. Ramanujam** *is with the Department of Electronics and Communication Engineering, Krishna Engineering College, Mohan nagar Ghaziabad, U.P- India.*

However, the Federal Trade Commission (FTC) [1,3]- the antitrust enforcement agency - is the organization leading the effort for resolution.

The contents of the paper is arranged a follows, in section-2 the levels of network interdependencies has been discussed, in section-3 different categories of networks have been discussed, packet blur deficiencies and technology developments are discussed in section 4, in section- 5 a new trend and section -6 transfor network architecture(TNA) and section- 7 different levels of software security attacks have been discussed, followed by planning of our future work in section-8, and at last we shall conclude the objectives of this work in section 9.

## 2 Network Interdependencies

Networks have become a tool for everyday social and business interactions, making it a part of the economy. Even though networking is technology-driven, many technical issues cannot be resolved in isolation in a modern economy. Technical issues involving networks now get intertwined with business, financial, social, and even political processes and systems. Factoring technology implications in business and economic decision-making has not kept pace with increased role of technology in the modern economy. Many problems in networking today can be traced to insufficient appreciation of these system interdependencies and deficiencies.

## 2.1 Denial of Responsibility

National Association for Security and Trust Evaluation warns of an increase in serious security breaches known as Denial of Responsibility (DoR) attacks. More recent DoR attacks include the inclusion of cool features" that benefit only a few curious experimenters but open the door to serious intrusions. DoR attacks are viral, in the sense that they begin in a governmental directive or software company, but spread rapidly to major customers who wish to minimize the risks created by the software flaws. Force the application to operate in low memory, and network-availability conditions. Executing an application, the computer loads it into memory and then gives the application additional memory to store and manipulate its internal data. Although, memory is temporary, it really be useful, an application needs to store persistent data. Without sufficient memory disk space, most applications will not perform their intended function. The objective of this attack is to deprive the application of any of these resources so testers can understand how robust and secure their application is under stress. The attacks caused the availability of the application as concern with the security aspects.

Independent entities. Lucent, which inherited most of the original Bell Labs, became a product organization retaining only market-driven research activities. The entities that emerged from the divestiture no longer had monopoly markets and were forced into organizations having financial objectives as their primary focus. Besides, the smaller revenue base was insufficient to maintain original level of investment in new technology innovation and scientific research.

## 2.2 Emergence of "Pecuniary entrepreneurship"

When an economy adopts large-scale innovations – whether technological or not - the result is a wave phenomenon. Best-known economic wave theory is called Kondratieff wave. A technology adoption wave has four phases:

1. Early adoption
2. Technology acceptance and growth
3. Financial market exuberance and bubble
4. Bubble collapse

The original economic wave phenomenon occurred during the "tulip Mania" in Holland in the 1630s. Under normal conditions, financial systems act as a facilitator for economic activity. During a large-scale technology adoption phase, financial markets enter a state of "irrational exuberance", following demonstrated practical benefits and value of new technologies. During this phase, financial markets try to drive the new technology adoption beyond the productive capabilities possible in the economy, creating "financial bubbles" i. However, this artificial growth cannot be sustained, resulting in the inevitable collapse.

The underlying dynamics can be understood from the "tulip mania". Tulips were introduced in Holland from Turkey, and became very popular as a symbol of wealth. The Dutch economy was already well developed and wealthy from trade.

A tulip bubble takes about six months to sprout, grow, and bloom -- an unbearably long time when there is a huge demand for it. Option trading in tulips was introduced to help wider participation – option contracts could be bought and sold on tulip bulbs until they bloom. The time it takes to complete an option trade is a tiny fraction of the six months it takes for a tulip bulb to become flower. This financial innovation enabled everyone in Holland who had interest, to participate in the "tulip market", trading as many times as they wished, driving up prices. The result was de-coupling of "tulip financial market" from "tulip production economy". Such a situation is unsustainable, as financial markets do not have an independent existence. Soon everybody in Amsterdam had "tulip contracts", as a result there was no new demand, and "tulip financial market" collapsed.

Building networks is a tedious process, taking several decades -- too long to suit the designs of financial markets. A similar exuberance developed for the Internet in the 1990s, creating the "Internet bubble". Even though the "Internet Bubble" collapsed, advances in "financial engineering" are helping continuation of market exuberance and creating a new "financial capitalism" -- triumph of the speculator over the manager and of the financier over the producerii. The current "liquidity bubble" in derivatives and hedge funds are a more elaborate application of the same principles used in "tulip markets"iii. Until financial markets reestablish equilibrium with sustainable development, economic benefits from advanced networking will have to wait.

## 2.3 Outdated Network Reference Model

For a variety of reasons, packet switching technology evolved in the 1960s and the 1970s in opposition to the prevailing Public Switched Telephone Networks (PSTN)iv. As a result, many key networking principles that were learned by trial and error since the first telephone call in 1876 were left out of packet networks. The Internet, the packet network that gained widespread adoption, also inherits many of these deficiencies.

As the Internet gained popularity, it was necessary to have a reference model to help direct further development of networks. The reference model for networks that came to be accepted is shown in Fig. 1, referred hereafter as the "Packet Cloud".

The first telephones had no network but were in private use, wired together in pairs. Users who wanted to talk to different people had as many telephones as necessary for the purpose. A user who wished to speak, whis-tled into the transmitter until the other party heard. Soon, however, a bell was added for signaling, and then a switch hook and telephones took advantage of the exchange principle already employed in telegraph networks. Each telephone was wired to a local telephone exchange, and the exchanges were wired together with trunks. Networks were connected together in a hierarchical manner until they spanned cities, countries, continents and oceans. This was the beginning of the PSTN, though the term was unknown for many decades.

Automation introduced pulse dialing between the phone and the exchange, and then among exchanges, followed by more sophisticated address signaling including multi-frequency, culminating in the SS7 network that connected most exchanges by the end of the 20th cent.

### 2.4 Packet-centric network reference model

There were many technical reasons for use the Packet Cloud as a reference during the initial growth phase of the Internet in the 1980s. When key packet technology developments were contemplated in the 1960s through the 1980s bandwidth was at a premium. The line speed in most early Internet links was 50 Kpbs. Under those line speeds, maximizing bandwidth utilization was of paramount importance. Packet networks offered superior bandwidth utilization compared to circuit switched networking, in which dedicated bandwidth allocation methods were used -- irrespective of actual use. There were no technologies in sight that could overcome this "bandwidth bottleneck".

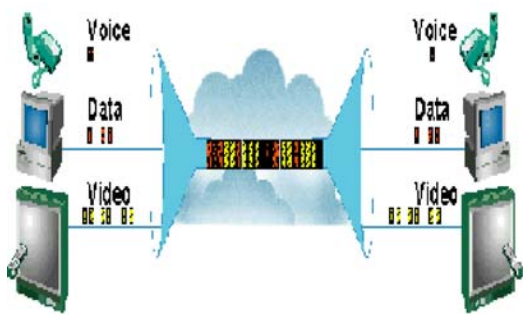

Fig. 1 "Packet Cloud", packet-centric network reference model

Soon the use of the Packet Cloud model became implicit and subconscious in the minds of network designers. The downsizing of the Bell Labs, which was occurring over the same period, reinforced this "Groupthink"v. New developments in laser and fiber optics technologies overcame the bandwidth bottleneck by the mid-1990s. However, the network industry has been unwilling to let go of the Packet Cloud dogmavi to take advantage of abundance of bandwidth and design better network systems.

## 3 NETWORK

Networks are often classified as Local Area Network **(LAN)**, Wide Area Network **(WAN)**, Metropolitan Area Network **(MAN)**, Personal Area Network **(PAN)**, Virtual Private Network **(VPN)**, Campus Area Network **(CAN)**, Storage Area Network **(SAN)**, etc. depending on their scale, scope and purpose. Usage, trust levels and access rights often differ between these types of network - for example, LANs tend to be designed for internal use by an organization's internal systems and employees in individual physical locations (such as a building), while WANs may connect physically separate parts of an organization to each other and may include connections to third parties.

### 3.1 Local Area Network

A local Area Network (LAN) is a computer network covering a small physical area, like a home, office, or small group of buildings, such as a school, or an airport. Current wired LANs are most likely to be based on Ethernet technology, although new standards like ITU-T also provide a way to create a wired LAN using existing home wires (coaxial cables, phone lines and power lines)[2].

For example, a library may have a wired or wireless LAN for users to interconnect local devices (e.g., printers and servers) and to connect to the internet. On a wired LAN, PCs in the library are typically connected by category 5 (Cat5) cable, running the IEEE 802.3 protocol through a system of interconnected devices and eventually connect to the Internet. The cables to the servers are typically on Cat 5e enhanced cable, which will support IEEE 802.3 at 1 Gbit/s. A wireless LAN may exist using a different IEEE protocol, 802.11b, 802.11g or possibly 802.11n. The staff computers (bright green in the figure) can get to the color printer, checkout records, and the academic network and the Internet. All user computers can get to the Internet and the card catalog. Each workgroup can get to its local printer. Note that the printers are not accessible from outside their workgroup.

### 3.2 Campus area network

A campus area network (CAN) is a computer network made up of an interconnection of local area networks (LANs) within a limited geographical area. It can be considered one form of a metropolitan area network, specific to an academic setting.

In the case of a university campus-based campus area network, the network is likely to link a variety of campus buildings including; academic departments, the university library and student residence halls. A campus area network is larger than a local area network but smaller than a wide area network (WAN) (in some cases).

The main aim of a campus area network is to facilitate students accessing internet and university resources. This is a network that connects two or more LANs but that is limited to a specific and contiguous geographical area such as a college campus, industrial complex, office building, or a military base. A CAN may be considered a type of MAN (metropolitan area network), but is generally limited to a smaller area than a typical MAN. This term is most often used to discuss the implementation of networks for a contiguous area. This should not be confused with a Controller Area Network. A LAN connects network devices over a relatively short distance. A networked office building, school, or home usually contains a single LAN, though sometimes one building will contain a few small LANs (perhaps one per room), and occasionally a LAN will span a group of nearby buildings.

### 3.3 Metropolitan area network

A metropolitan area network (MAN) is a network that connects two or more local area networks or campus area networks together but does not extend beyond the boundaries of the immediate town/city. Routers, switches and hubs are connected to create a metropolitan area network.

### 3.4 Wide area network

A wide area network (WAN) is a computer network that covers a broad area (i.e. any network whose communications links cross metropolitan, regional, or national boundaries [1]). Less formally, a WAN is a network that uses routers and public communications links. Contrast with personal area networks (PANs), local area networks (LANs), campus area networks (CANs), or metropolitan area networks (MANs), which are usually limited to a room, building, campus or specific metropolitan area (e.g., a city) respectively. The largest and most well-known example of a WAN is the Internet. A WAN is a data communications network that covers a relatively broad geographic area (i.e. one city to another and one country to another country) and that often uses transmission facilities provided by common carriers, such as telephone companies. WAN technologies generally function at the lower three layers of the OSI reference model: the physical layer, the data link layer, and the network layer.

### 3.5 Global area network

A global area networks (GAN) specification is in development by several groups, and there is no common definition. In general, however, a GAN is a model for supporting mobile communications across an arbitrary number of wireless LANs, satellite coverage areas, etc. The key challenge in mobile communications is "handing off" the user communications from one local coverage area to the next. In IEEE Project 802, this involves a succession of terrestrial WIRELESS local area networks (WLAN).

### 3.6 Virtual private network

A virtual private network (VPN) is a computer network in which some of the links between nodes are carried by open connections or virtual circuits in some larger network (e.g., the Internet) instead of by physical wires. The data link layer protocols of the virtual network are said to be tunneled through the larger network when this is the case. One common application is secure communications through the public Internet, but a VPN need not have explicit security features, such as authentication or content encryption. VPNs, for example, can be used to separate the traffic of different user communities over an underlying network with strong security features.

A VPN may have best-effort performance, or may have a defined service level agreement (SLA) between the VPN customer and the VPN service provider. Generally, a VPN has a topology more complex than point-to-point.

A VPN allows computer users to appear to be editing from an IP address location other than the one which connects the actual computer to the Internet.

### 3.7 Internetwork

An Internetwork is the connection of two or more distinct computer networks or network segments via a common routing technology. The result is called an internetwork (often shortened to internet). Two or more networks or network segments connect using devices that operate at layer 3 (the 'network' layer) of the OSI Basic Reference Model, such as a router. Any interconnection among or between public, private, commercial, industrial, or governmental networks may also be defined as an internetwork.

In modern practice, interconnected networks use the Internet Protocol. There are at least three variants of internetworks, depending on who administers and who participates in them: Intranet Extranet Internet

Intranets and extranets may or may not have connections to the Internet. If connected to the Internet, the intranet or extranet is normally protected from being accessed from the Internet without proper authorization. The Internet is not considered to be a part of the intranet or extranet, although it may serve as a portal for access to portions of an extranet.

### 3. 8 Intranet

An intranet is a set of networks, using the Internet Protocol and IP-based tools such as web browsers and file transfer applications, that is under the control of a single administrative entity. That administrative entity closes the intranet to all but specific, authorized users. Most commonly, an intranet is the internal network of an organization. A large intranet will typically have at least one web server to provide users with organizational information.

### 3.9 Extranet

An extranet is a network or internetwork that is limited in scope to a single organization or entity and also has limited connections to the networks of one or more other usually, but not necessarily, trusted organizations or entities (e.g., a company's customers may be given access to some part of its intranet creating in this way an extranet, while at the same time the customers may not be considered 'trusted' from a security standpoint). Technically, an extranet may also be categorized as a CAN, MAN, WAN, or other type of network, although, by definition, an ex-tranet cannot consist of a single LAN; it must have at least one connection with an external network.

### 3.10 Internet

The Internet consists of a worldwide interconnection of governmental, academic, public, and private networks based upon the networking technologies of the Internet Protocol Suite. It is the successor of the Advanced Research Projects Agency Network (ARPANET) developed by DARPA of the U.S. Department of Defense. The Internet is also the communications backbone underlying the World Wide Web (WWW). The 'Internet' is most commonly spelled with a capital 'I' as a proper noun, for historical reasons and to distinguish it from other generic internetworks.

Participants in the Internet use a diverse array of methods of several hundred documented, and often standardized, protocols compatible with the Internet Protocol Suite and an addressing system (IP Addresses) administered by the Internet Assigned Numbers Authority and address registries. Service providers and large enterprises exchange information about the reach ability of their address spaces through the Border Gateway Protocol (BGP), forming a redundant worldwide mesh of transmission paths.

## 4 PACKET BLUR DEFICIENCIES

The strength of the Packet Cloud reference model is its simplicity. Use packet switching for any type of traffic: voice, data or video. There would be only "one network" to maintain and manage, if this model were adopted. Another potential benefit claimed is reuse of designs, when everything is packet-based. But the arguments in favor of Packet Cloud reference model fail to take into account its deficiencies and alternate choices available for creating better network designs.

One of the glaring gaps in the Packet Cloud model is lack of out-of-band signaling. Keeping networks operating properly require "control packets", data about status of networks and devices, as well as commands to and from devices and systems attached to networks. One of the efficiencies in packet switching is gained by "treating all packets equally", including control packets. But during network congestion or other problems, control packets sent to resolve problems get delayed or lost creating a vicious circle of degraded network performance and functionality. Such problems do not happen in the PSTN because of the highly reliable Signaling System 7 (SS7). In the PSTN, control data and user data (datapath) have separate network links – they do not share the same links ("out-of-band").

Congestion in datapath networks does not degrade the SS7 network as it is designed with over capacity, redundancy and maximum reliability to function even in worst-case failures. Denial-of-Service (DoS) and Distributed-DoS (D-DoS) in the Internet are a byproduct of lack of an out-of-band control network, similar to the SS7 for the PSTNvii. However, adding such out-of-band signaling is inconsistent with packet network design ideals.

### 4.1 Technological development

In common with most countries, the development of technology allowed for different networking, and the maintenance of a formal hierarchy disappeared into a distributed network. By the mid-1990s, a revised structure had appeared, reflected by the replacement of the old departmental area codes by the assignment of regional codes and a major renumbering scheme for strategic planning, privatization, and deregulation under the auspices of ART, the Autorité de régulation des télécommunications (Regulatory Authority for Telecommunications - since 2005, ARCEP, as responsibility for postal services was added). After 1996, the country prepared for complete deregulation of the telephone network.

Thus, the local exchanges (zones à autonomie d'acheminement) are connected somewhat differently by various carriers. However, the largest of these, based upon the (partially) privatized former government network, is a two-level long distance hierarchy, based on 80 CTS (centre de transit secondary) and 8 CTP (centre de transit primaries) locations. In addition, there are 12 CTI (centre de transit internationaux) for connections to areas which are not integrated into the French telephone network [note that some overseas locations are considered "domestic" for telecommunications purposes].

When packet networks were developed, it was designed to carry computer data that tolerate transit delays. "Best effort routing" is a key principle of packet switching permitting delay and loss of packets while keeping the network operating for delay-tolerant traffic. However, success of the Internet induced packet-enthusiasts to overreach the design goals of the Internet, designing Voice-over-IP (VoIP) to counter overbearing telcos. Since telephone voice is real-time (propagation delay needs to be less than 200 ms for toll-grade voice), using the Internet for such traffic is contradictory to its design constraints. This contradiction has been re-framed as the "Quality of Service" (QoS) problem. Solutions developed for achieving desired QoS produce other side effects. The side effects of high QoS solutions degrade the original design principles of the Internet. The Internet, having become a public medium, generates public awareness of the issues -- turning technology issues into political controversies.

## 5 A NEW TREND

Inconsistency with existing network infrastructure is another deficiency of the Packet Cloud. Except for new all-packet segments, the Internet is an overlay over the preexisting PSTN, based on time-division-multiplexing systems (TDM). These dedicated, channel-allocating TDM systems do not fit the Packet Cloud model. Therefore, acceptance of the Packet Cloud model automatically obsoletes the TDM infrastructure, which is huge, costing hundreds of billions or even trillions of dollars, based on the valuation methodology used. This premature obsolescence of useful infrastructure is one of the critical reasons for failure and lack of success of packet-centric implementations.

New solutions and approaches for solving network problems need to accept certain constraints. One approach often suggested is "fork-lift solutions", i.e., throwing away everything and starting overviii. Such solutions were possible in the early stages of the Internet development, when usage was confined to universities and research labs. Now, the Internet has become an everyday tool for social and business transactions. Expecting all Internet users to switch over to some new way in a prescribed manner and timeframe is doomed to fail. Only solutions that can reach majority usage voluntarily are likely to succeed – solutions that offer added value and benefits to users by switching to new solutions1ix. Final steps of a full migration to a new technology generation may be implemented in a gradual manner, like the turning-off of the over-the-air TV transmission in the USx. Such practices are normal in other economic sectors. For example, in civil engineering, roads and bridges are routinely upgraded by building a bypass while new construction is in progress.

## 6 THE TRANSFER NETWORK ARCHITECTURE (TNA)

Network Architecture (TNA) is a new model created by superimposing an architecture framework to the existing network infrastructure; consisting of the Internet, Signaling System 7 (SS7) network, data circuits of the PSTN, and other networks (Fig 2). This super architecture is made possible with the introduction of a new network, the Access Network – to provide connectivity and transport functions between customer premise systems and backbone network systems.

The key advantage of the TNA model is that it represents the current state of networks closely (compared to Packet Cloud, Fig. 1), and thus has more practical value for using in design and deployment of next generation systems. Underlying the new model is the assumption that a revolutionary approach of replacing all existing systems with new systems is not viable, due to the costs and operational constraints. The TNA model, in contrast, provides an evolutionary approach that can coexist with current systems, and provides for gradual migration to superior solutions, constructed with best-of-breed heterogeneous systems. The TNA model permits the use of both packet and circuit switching technologies for designing and deploying network systems, and allows for interconnecting different networks using all possible combinations of network technologies.

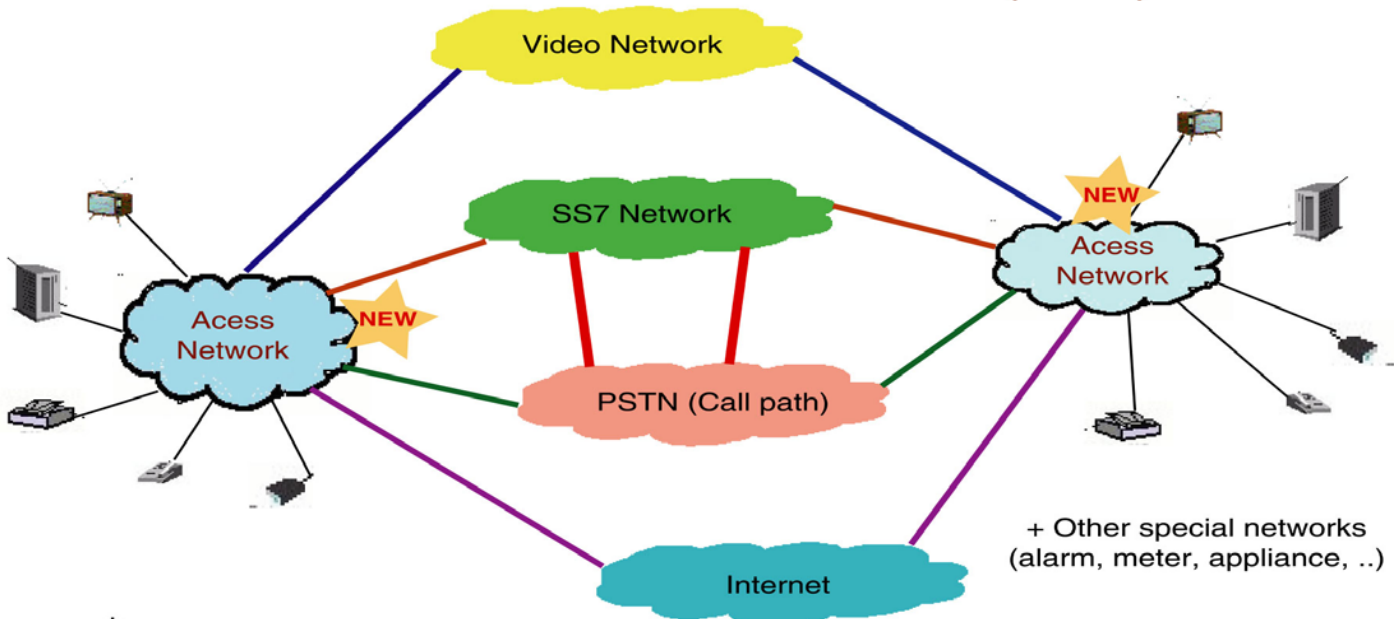

Many current and new products that are offering local access solutions using various technologies (like VoIP, xDSL) are implemented in a manner to make Access Networks part of the Internet, and completely eliminate PSTN in the future. TNA is a reference model for designing networks with backward and forward compatibility with TDM networks and products, without making it necessary to obsolete useful systems.

In the TNA model, backbone networks can be designed to offer specialized capabilities for different types of traffic: voice, data, video, or special services (such as utility meters, surveillance devices, alarm monitors, appliance networks, etc.) The Access Networks transport and transfer network traffic between subscriber devices and backbone networks, and implement connectivity functions. Network devices within each of these networks provide specialized capabilities optimized for the services offered by the networks they belong to.

The TNA model provides a natural path for phased network evolution by providing compatibility with existing network infrastructure and products, while protecting current infrastructure, product, operation management systems, and human training investments. In addition, the TNA model allows flexible network design and deployment strategies with maximum flexibility — without mandating either a packet-centric or a circuit-centric approach. The design choices are left to be decided based on service, performance, and cost considerations.

The TNA model helps in making macro-level decisions about network systems, and a framework for defining the functionality of network products based on its role in a sub-network. The primary barrier for network evolution to meet many current future needs is the lack of viable Access Networks, with the type functional capabilities defined in the TNA model.

## 7. CLASSIFICATION OF ATTACKS

Software security attacks are classified according to the problem when these events will be realized (Kinds of attack), the attack kinds are of the following kinds:

### 7.1. Disclosure

Disclosure is the security problem caused when an attacker can acquire the information of a system in unauthorized way.

### 7.2. Deception

Deception is the way through which the system accepts the false information entered by an unauthorized user to retrieve the important information.

### 7.3 Disruption

Disruption is the security problem caused when an attacker can interrupt or prevent correct operation of the system.

### 7.4 Usurpation

Attackers cause usurpation when they can control some part of a system for misuse in unauthorized way.

## 8 FUTURE WORK

In this paper we tried to give the formal way to prevent New Model of Network. A formal approach of New Model of Network always considered as the best solution to protect corporate resources with fewer efforts. However, little work has been done in this aspect, though we are trying to provide the techniques to alleviate the above Network attacks, which will provide best ROI to organizations that integrate it as part of their Network development and Security attacks.

## 9 CONCLUSION

In this work, we have described some of the previous efforts to measure SS7, PSTN, Network, and we have outlined some of the difficulties that have been encountered. We believe that a periodic, comprehensive evaluation of TNA could be valuable for network managers, information security officers and data managers. However, The Transfer Network Architecture (TNA) is a reference model that can help such developments. For many reasons, including those discussed herein, network industry is presently unable to make technology choices based on technical merit; instead focus on turf battles and petty politics – resulting in the current stagnation. The current packet-centric strategies are not helpful to meet the network needs of today and tomorrow. Superior network capabilities will emerge by promoting technology and protocol agnostic inter-operating networks, with each network implemented by optimizing its design requirements derived from service needs, without extraneous considerations. The Transfer Network Architecture (TNA) is a reference model that can help such developments.

## ACKNOWLEDGMENT

The Success of this research work would have been uncertain without the help and guidance of a dedicated group of people. I would like to express my true and sincere acknowledgements as the appreciation for their contributions. I also express my sincere thanks to Dr. Ajay Sharma, Director General, KIET, Ghaziabad, India and Prof. Verghese Mundamattam, Director, Prof. BPG Raju, Principal, Prof. George Varughese, Advisor, ISBAT, Kampala, Uganda, East Africa, for their encouragement and support. Last but not least I hearltly thank to my family members my father Prof. D.N.Singh, wife Nirmala Singh, Daughter Kanchan Singh for their cooperation.## ACKNOWLEDGMENT

The Success of this research work would have been uncertain without the help and guidance of a dedicated group of people. I would like to express my true and sincere acknowledgements as the appreciation for their contributions. I also express my sincere thanks to Dr. Ajay Sharma, Director General, KIET, Ghaziabad, India and Prof. Verghese Mundamattam, Director, Prof. BPG Raju, Principal, Prof. George Varughese, Advisor, ISBAT, Kampala, Uganda, East Africa, for their encouragement and support. Last but not least I hearltly thank to my family members my father Prof. D.N.Singh, wife Nirmala Singh, Daughter Kanchan Singh for their cooperation.

**Ram Kumar Singh** [ M.Tech.(Comp. Sci. & Engg.), B.Engg. (Electronics and Communication Engg.), CCNA, Polyt Echnichnic Diploma in Computer Science] is Senior Lecturer and CISSP Trainer in Department of Information Technology, International School of Business and Technology, Kampala,Uganda. His research area is Network Security and Information Security.

**Prof. T. Ramanujam** is the (Founder) Director of Krishna Institute of Engineering and Techology, Mohan Nagar, Ghaziabad, U.P., India. He is currently doing Ph D in areas related to computational Biology and Bioinformatics, M Tech (Electrical Engg with specialisation in Computer Science from IIT, Kanpur) and M Sc in Military History from Madras University, BE(Hons) in Electrical Engg, . He retired as an Air Vice Marshal of Indian Air Force after 35 years of distinguished service and is in the field of academics for the last 8 years. He was Dean (Academics), KIET, Ghaziabad for 2 years.